\documentclass[letterpaper,conference]{IEEEtran}
\pdfoutput=1

\usepackage{amsmath,amssymb,eucal,graphicx,color}
\usepackage{pst-plot}
\usepackage{pst-node}
\usepackage{epsfig}
\usepackage{amsthm}
\usepackage{amsfonts}
\usepackage{epsfig}
\usepackage{float}
\usepackage{pstricks}
\usepackage{color}
\usepackage{multicol}
\usepackage{subfigure}

\newtheorem{theorem}{Theorem}

\begin{document}
\title{An Outer Bound for the Memoryless Two-user Interference Channel with General Cooperation}

\author{%
\IEEEauthorblockN{Daniela Tuninetti}
\IEEEauthorblockA{%
University of Illinois at Chicago,\\
Department of Electrical and Computer Engineering\\
Chicago, IL 60607, USA\\
Email: danielat@uic.edu}
}
\maketitle

\begin{abstract}
The interference channel models a wireless network where several source-destination pairs compete for the same resources. When nodes transmit simultaneously the destinations experience interference. This paper considers  a 4-node network, where two nodes are sources and the other two are destinations. All nodes are full-duplex and cooperate to mitigate interference. A sum-rate outer bound is derived, which is shown to unify a number of previously derived outer bounds for special cases of cooperation. The approach is shown to extend to cooperative interference networks with more than two source-destination pairs and for any partial sum-rate. How the derived bound relates to similar bounds for channel models including cognitive nodes, i.e., nodes that have non-causal knowledge of the messages of some other node, is also discussed. Finally, the bound is evaluated for the Gaussian noise channel and used to compare different modes of cooperation.
\end{abstract}


\section{Introduction}
\label{sec:intro}
Understanding how to manage interference in wireless networks has been an area of intense work over the past few years. Although the exact capacity characterization of the simplest interference channel model with two sources and two destinations sharing the same channel is still open in general, progress has been made for the Gaussian noise case. In the seminal paper by Etkin et al~\cite{etw} the capacity region of the 2-user Gaussian interference channel has been universally characterized to within 1~bit for all channel parameters.  Following the approach of~\cite{etw}, several works investigated how cooperation can improve rate performance with respect to the classical non-cooperative case. 

Host-Madsen~\cite{HostMadsen} considered the Gaussian noise channel with either source or destination cooperation; he developed inner and outer bounds and showed that cooperation does not increase the degrees of freedom (DoF) of the channel. 
The case of source cooperation/generalized feedback has been investigated in~\cite{tuniita2010,TuninettiEchoJ1,echoasilomar2011,VinodTX}; in particular~\cite{TuninettiEchoJ1} has the largest known inner bound,~\cite{VinodTX} characterized the Gaussian sum-capacity to within 19~bits (lately improved to 2~bits in `strong cooperation' by~\cite{echoasilomar2011}), and~\cite{tuniita2010} proposed a framework to determine sum-rate upper bounds for general source cooperation; source cooperation includes as a special case conferencing encoders~\cite{ihunagtseTX} and output feedback~\cite{SuhTseOutputFeedback,melda}.

The case of destination cooperation in Gaussian noise has been investigated in~\cite{ihunagtseRX} for the out-of-band case and in~\cite{VinodRX} for the in-band case, where the capacity has been determined to within a constant gap.

The case of general cooperation, i.e., where all sources and all destinations cooperate, has not been studied in full generality. In~\cite{VahidRateLimitedFeedback} the symmetric capacity for the Gaussian noise channel with out-of-band rate-limited feedback from a source to its intended destination has been determined to within a constant gap. Recently, the work in~\cite{VahidRateLimitedFeedback} has been extended in~\cite{suhISIT2012} so as to model the feedback channel as a deterministic/noiseless interference channel. In this model the sources can communicate with the destinations and the destinations with the sources, but intra source or intra destination communication is not allowed.

Most work focused on Gaussian noise channels where inner bounds are based on rate-splitting, superposition coding and binning and where the approximately optimal rate and power splits are usually inspired by the analysis of the linear deterministic approximation of the Gaussian channel at high SNR. Outer bounds are more of an art with every work proposing--in addition to the classical cut-set bound--an ad-hoc generalization of the bounds for the classical non-cooperative interference channel~\cite{etw}. In this work we seek to derive an outer bound for the case of general cooperation on a memoryless interference channel.

\smallskip
Our main contribution is a generalization of the outer bound of~\cite{tuniita2010}--developed for the case of source cooperation and inspired by~\cite[Thm.1]{kramer}--to the case of general cooperation. We show that the new bound recovers as special cases all known bounds that reduce to~\cite[Thm.1]{kramer} in case of no-cooperation. We also show that the new bound continues to hold for a more general class of channels, such as cognitive channels, thereby explaining why outer bounds for different channel models behave similarly for a certain range of parameters. We evaluate the bound for the Gaussian noise channel and use it to compare different modes of cooperation.

\smallskip
The rest of the paper is organized as follows. 
Section~\ref{sec:chmodel} introduces a general model for cooperation on the interference channel.
Section~\ref{sec:outer} proves the main result of the paper and discusses its relationship with known results.
Section~\ref{sec:awgn} evaluates the bound for the Gaussian noise channel.
Section~\ref{sec:conc} concludes the paper.

\section{Channel Model}
\label{sec:chmodel}


{\bf Two-way Interference Channel.}
Consider a network of $2K$ {\em full-duplex} nodes, 
with input alphabets $\mathcal{X}_{1},\ldots,\mathcal{X}_{2K}$, 
output alphabets $\mathcal{Y}_{1},\ldots,\mathcal{Y}_{2K}$, and 
a channel transition probability $P_{Y_{1},\ldots,Y_{2K}|X_{1},\ldots,X_{2K}}(y_{1},\ldots,y_{2K}|x_{1},\ldots,x_{2K}) : \mathcal{Y}_{1}\times\ldots\times\mathcal{Y}_{2K} \to [0,1]$ for all $(x_{1},\ldots,x_{2K})\in \mathcal{X}_{1}\times\ldots\times\mathcal{X}_{2K}$.
Node $i$, $i\in[1:2K]$
it has an independent message $W_i$ to send,
it has input to the channel $X_i\in\mathcal{X}_i$,
it has output from the channel $Y_i\in\mathcal{Y}_i$, and
it is interested in decoding message $W_{i+K}$.
Since node $i\in[1:2K]$ sends a message to node $i+K$ and receives a message from node $i+K$, we say that nodes $i$ and $i+K$ form a {\em two-way pair}. The network is composed of $K$ such two-way pairs sharing the same physical channel. 
The channel is {\em memoryless}, i.e., the following Markov chain holds
\begin{align}
&(W_{1},\ldots,W_{2K},Y_{1}^{t-1},\ldots,Y_{2K}^{t-1},X_{1}^{t-1},\ldots,X_{2K}^{t-1})
\nonumber\\
&\to (X_{1,t},\ldots,X_{2K,t})
\to (Y_{1,t},\ldots,Y_{2K,t}),
\ t\in\mathbb{N}_+,
\label{eq:memoryless channel}
\end{align}
where $A^{t}$ denotes a vector of length $t$ with components $(A_{1},\ldots,A_{t})$
and where by convention $A^{0}=\emptyset$.

A non-negative rate vector $(R_{1},\ldots,R_{2K})$ is said to be $\epsilon$-achievable, for some $\epsilon\in(0,1)$, if there exists a family of length-$N$ $(N,2^{N R_1},\ldots,2^{N R_{2K}})$-codes consisting of 
encoding functions $\mathsf{e}^{(N)}_{1},\ldots,\mathsf{e}^{(N)}_{2K}$, where
\[
\mathsf{e}^{(N)}_{i} : [1:2^{N R_{i}}]\times \mathcal{Y}_i^{t-1} \to \mathcal{X}_i
\]
such that at time $t\in[1:N]$ the channel input of node $i\in[1:2K]$ is determined by
$
X_{i,t}:=\mathsf{e}^{(N)}_{i}(W_{i},Y_i^{t-1});
$
decoding functions $\mathsf{d}^{(N)}_{1},\ldots,\mathsf{d}^{(N)}_{2K}$, where
\[
\mathsf{d}^{(N)}_{i} :  [1:2^{N R_{i}}]\times \mathcal{Y}_i^{N} \to [1:2^{N R_{i+K}}]
\]
such that at time $t=N$ the message estimate of node $i\in[1:2K]$ is determined by
$
\widehat{W}_{i+K} :=\mathsf{d}^{(N)}_{i}(W_{i},Y_i^N);
$
and such that the probability of error satisfies
\[
\max_{i\in[1:2K]}\mathbb{P}[\widehat{W}_{i+K}\not= W_{i}] \leq \epsilon.
\]
The capacity region is the closure of the set of the rate vectors $(R_{1},\ldots,R_{2K})$ that are $\epsilon$-achievable for all $\epsilon\in(0,1)$~\cite{book:cover_thomas:it}.

\begin{figure}
\centering
%
%
\includegraphics[width=7cm]{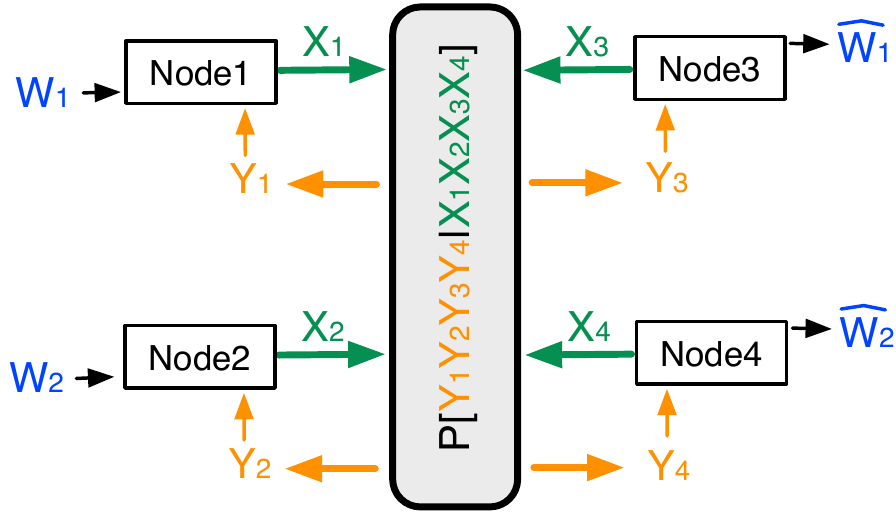}
\caption{A two-user general cooperative interference channel.}
\label{fig:gen coop 2user ifc}
\end{figure}

\smallskip
{\bf General Cooperative Interference Channels.}
In this work we focus on the capacity region of the following class of two-way interference networks.
A two-way interference channel is said to be a general memoryless {\em cooperative $K$-user interference channel} ($K$-CoopIFC)
if ``messages only flow in one direction'', that is to say, if $R_{K+1}=\ldots=R_{2K}=0$. 
In this case, nodes $1$ to $K$ are sources and nodes $K+1$ to $2K$ are destinations. 
Fig~\ref{fig:gen coop 2user ifc} shows a $K$-CoopIFC for $K=2$.

%

\smallskip
{\bf Types of Cooperation.}\label{sec:types of coop}
Several cooperation models have been analyzed in literature.
Cooperation can be {\em in-band} or {\em out-of-band}. For out-of-band cooperation, the network is effectively composed of two parallel networks: the underlying interference channel (from the source inputs to the destination outputs) and the cooperation channel (which is usually assumed to be deterministic/noiseless). 
We further distinguish:

{\em Source Cooperation or Generalized Feedback.}
The destinations do not have an input to the channel, i.e.,
\[
X_{K} = \ldots = X_{2K-1} = \emptyset.
\]
This includes as special cases:
output feedback from a source to a destination~\cite{SuhTseOutputFeedback,melda,TadonUlukus},
conferencing encoders~\cite{ihunagtseTX}, and
in-band source cooperation~\cite{tuniita2010,VinodTX}.

{\em Destination Cooperation.}
The sources do not have an output from the channel, i.e.,
\[
Y_{1} = \ldots = Y_{K} = \emptyset. 
\]
This includes as special cases:
conferencing decoders~\cite{ihunagtseRX} and
in-band destination cooperation~\cite{VinodRX}.

{\em General Cooperation.}
This is the most general case where all nodes cooperate.
This includes as special cases, besides the cases of
source and destination cooperation mentioned above,
out-of-band rate-limited feedback from a source to the intended destination~\cite{VahidRateLimitedFeedback} and
the two-way-like model considered in~\cite{suhISIT2012}.

{\em Ultimate Limit of Cooperation.}
By sharing the message vector $(W_{1},\ldots,W_{K})$
among the sources and the channel output vector $Y_\mathrm{eq}:=(Y_{K+1},\ldots,Y_{2K})$
among the destinations
one obtaines an equivalent memoryless point-to-point channel with
input  $X_\mathrm{eq} := (X_{1},\ldots,X_{K})$ and
output $Y_\mathrm{eq}$
whose capacity gives the following sum-rate upper bound
\[
R_{1}+\ldots+R_{K} \leq  \max_{\mathbb{P}[X_\mathrm{eq}]} I(X_\mathrm{eq}; Y_\mathrm{eq}),
\]
which cannot be further improved by the availability of the feedback
$(Y_{1},\ldots,Y_{K})$~\cite{book:cover_thomas:it}.

\medskip
{\bf The Gaussian Noise Channel.}
To make the above more concrete, consider the Gaussian noise channel.
We change slightly the notation for this section  and
indicate with $X$ a channel input in the underlying interference channel 
and with $V$ all other inputs. The complex-valued Gaussian $K$-CoopIFC has input-output relationship 
\begin{align}
Y_\ell &= \begin{bmatrix}
   f_\ell(V_{1},\ldots,V_{2K})\\
   \underbrace{\sum_{i=1  }^{ K} \mathrm{h}_{\ell,i} X_i}_{\text{from the sources}}
 + \underbrace{\sum_{i=K+1}^{2K} \mathrm{h}_{\ell,i} X_i}_{\text{from the destinations}} + Z_\ell
   \end{bmatrix}, \ \ell\in[1:2K],
\label{eq:awgnch}
\end{align}
where we assume:
(a) the channel gain matrix $\mathbf{H} := [\mathrm{h}_{\ell,i}]_{(\ell,i)\in[1:2K]\times[1:2K]}$ is constant and therefore known to all nodes,
(b) the input $X_i$ is subject to the average power constraint $\mathbb{E}[|X_i|^2] \leq \mathsf{P}_i$, $i\in[1:2K]$,
(c) the noise vector $\mathbf{Z} := [Z_{1},\ldots,Z_{2K}]$ is proper-complex with zero mean and covariance matrix $\Sigma_{\mathbf{Z}} \succ \mathbf{0}$ 
(without loss of generality we can set the diagonal entries of $\Sigma_{\mathbf{Z}}$ to one; depending on the type of cooperation, certain off-diagonal entries of $\Sigma_{\mathbf{Z}}$ do not affect the capacity region and can be chosen so as to tighten the outer bound~\cite{tuniita2010}), and
(d) the deterministic and discrete-valued function $f_\ell$ takes at most $2^{\mathsf{C}_\ell}$ values for some $\mathsf{C}_\ell\geq 0$, $\ell\in[1:2K]$. 
Without loss of generality one can set $\mathrm{h}_{i,i}=0$ because a node can always subtract its input $X_i$ from its channel output $Y_i$, $i\in[1:2K]$.

For $K=2$, the  {\em  baseline} to compare the gains of cooperation is the classical two-user interference channel with
\begin{align}
Y_\ell &= 0, \ \ell\in[1:2], \nonumber \\
Y_\ell &= \mathrm{h}_{\ell,1} X_1 + \mathrm{h}_{\ell,2} X_2 + Z_\ell, \ \ell\in[3:4], \label{eq:plainIFC}
\end{align}
whose capacity (exact or to within one bit) is discussed in~\cite{etw}.

The different types of cooperation
are obtained from~\eqref{eq:awgnch} by imposing the conditions  discussed previously.
For example, in-band cooperation is obtained with $f_\ell = 0$ for all $\ell\in[1:2K]$, while
out-of-band cooperation by setting $\mathrm{h}_{\ell,i}=0$ for either $\ell\in[1:K]$ or $i\in[K:2K-1]$ with $K=2$.
The {\em ultimate limit of cooperation} is
$R_1+R_2
\leq 
\mathsf{C}_{\rm MIMO}      
+\mathsf{C}_3+\mathsf{C}_4, 
$ 
where $\mathsf{C}_{\rm MIMO}$ is the capacity of the MIMO channel with channel matrix obtained from~\eqref{eq:plainIFC}
with perfect source and destination cooperation
and with per-antenna power constrain.

\section{Main Result: An Outer Bound for the General Memoryless $2$-CoopIFC}
\label{sec:outer}

Before presenting our main result, we remind the reader that the cut-set outer bound~\cite{book:cover_thomas:it} applied to a general network with independent messages at each node states that an achievable rate vector must satisfy
\begin{align}
R(\mathcal{S}\to\mathcal{S}^c) \leq I(X(\mathcal{S}); Y(\mathcal{S}^c)|X(\mathcal{S}^c))
\label{eq:general cuset}
\end{align}
for some joint distribution on the inputs,
where $\mathcal{S}$ is a subset of the set of all nodes in the network,  $\mathcal{S}^c$ is the complement of $\mathcal{S}$, and $R(\mathcal{S}\to\mathcal{S}^c)$ indicates the sum of the rates from the source nodes in $\mathcal{S}$ to the destination nodes in $\mathcal{S}^c$. For the $2$-CoopIFC:
\begin{theorem}\label{theorem:out cutset}
For the $2$-CoopIFC, an achievable rate pair $(R_1,R_2)\in\mathbb{R}^2_+$ must satisfy for some input distribution $\mathbb{P}[X_1,X_2,X_3,X_4]$ the following constraints
\begin{subequations}
\begin{align}
\hline
&\mathcal{S}, \ 
\mathcal{S}^c 
&& \text{\rm rate bound from~\eqref{eq:general cuset}} \nonumber
\\
\hline
&\{  1,2,4\} 
, \ \{  3\} 
&& R_1            \leq  I(X_1,X_2,X_4; Y_3|X_3), \label{eq:cutset r1 a}
\\
&\{  1,4\}   
, \ \{  2,3\}   
&& R_1            \leq  I(X_1,X_4; Y_2,Y_3|X_2,X_3), \label{eq:cutset r1 b}
\\
&\{  1\} 
, \ \{  2,4,3\} 
&& R_1            \leq  I(X_1; Y_2,Y_3,Y_4|X_2,X_3,X_4), \label{eq:cutset r1 c}
\\
\hline
&\{  2,1,3\} 
, \ \{  4\} 
&& R_2            \leq I(X_1,X_2,X_3; Y_4|X_4), \label{eq:cutset r2 a}
\\
&\{  2,3\} 
, \ \{  1,4\} 
&& R_2            \leq I(X_2,X_3; Y_1,Y_4|X_4,X_1), \label{eq:cutset r2 b}
\\
&\{  2\} 
, \ \{  1,3,4\} 
&& R_2            \leq I(X_2; Y_1,Y_3,Y_4|X_4,X_1,X_3), \label{eq:cutset r2 c}
\\
\hline
&\{  1,2\}   
, \ \{  4,3\}   
&& R_1+R_2        \leq I(X_1,X_2;Y_3,Y_4|X_4,X_3). \label{eq:cutset r1+r2}
\\\hline\nonumber
\end{align}
\label{eq:cuset}
\end{subequations}
\end{theorem}
\vspace*{-0.65cm}

The cut-set bound holds in great generality but it is known to be loose in general~\cite{book:cover_thomas:it}.
%
%
Our main result is the following sum-rate outer bound whose proof can be found in the Appendix:
\begin{theorem}
\label{theorem:out for general IFC-GF}
For the general 2-CoopIFC if a rate pair $(R_1,R_2)\in\mathbb{R}^2_+$
is achievable then, in addition to the cut-set bounds in~\eqref{eq:cuset}, it must satisfy
for some $\mathbb{P}[X_1,X_2,X_3,X_4]$
\begin{subequations}
{
\begin{align}
R_1+R_2  &\leq  I(X_1;Y_3,Y_2| Y_4,X_2,X_3,X_4) 
\nonumber\\&\quad
+ I(X_1,X_2,X_3;Y_4|X_4)\label{kramer-sr b},\\
R_1+R_2  &\leq  I(X_2;Y_4,Y_1| Y_3,X_1,X_3,X_4) 
\nonumber\\&\quad
+ I(X_1,X_2,X_4;Y_3|X_3)\label{kramer-sr a}. 
\end{align}
\label{eq:out for general IFC-GF}
}
\end{subequations}
\end{theorem}

\smallskip
{\bf Remark: Relationship between Thm.~\ref{theorem:out for general IFC-GF} and known results for special types of cooperation.}
The bounds in~\eqref{eq:out for general IFC-GF} generalize~\cite[Th.II.1]{tuniita2010}, which was derived for the case of general source cooperation/generalized feedback, to the case of general cooperation. Our contribution in this work is to show a single unifying way to derive all known bounds for cooperative IFCs that have appeared in the literature, including not only the case of source cooperation (which inspired Thm.~\ref{theorem:out for general IFC-GF}) but also destination and general cooperation.

To see how our bounds reduce to known bounds for the different special cases of cooperation listed in Section~\ref{sec:types of coop}, when applying the conditions that define each mode of cooperation, consider the following examples--the same extends to others types of cooperation. 

\smallskip
Example~1: In-band destination cooperation~\cite{VinodRX}.
With $Y_1=Y_2=\emptyset$ the sum-rate bound in~\eqref{kramer-sr a} reduces to
\begin{align*}
  &R_1+R_2
\\&\leq I(X_1,X_2,X_4;Y_3|X_3)+I(X_2;Y_4| Y_3,X_1,X_3,X_4)
\\&= H(Y_3|X_3)-H(Y_3| X_1,X_2,X_3,X_4)
\\&+ H(Y_4| Y_3,X_1,X_3,X_4)-H(Y_4| X_1,X_2,X_3,X_4, Y_3)
\end{align*}
The work in~\cite{VinodRX}, which was limited to additive noise channels only,
proved this upper bound in~\cite[page 208, first equation in the right column]{VinodRX}, where
$H(Y_\ell| X_1,X_2,X_3,X_4)$ is the entropy of the additive noise at node $\ell\in[3,4]$,
$H(Y_4| Y_3,X_1,X_3,X_4)$ is the entropy of the noisy observation of $X_2$ at node~4
conditioned on the noisy observation of $X_2$ at node~3, after all the other inputs have been removed, and
$H(Y_3|X_3)$ is the entropy of the channel output at node~3 after having removed the contribution of its
transmitted signal (in~\cite{VinodRX} the conditioning on $X_3$ is not present because $X_3$
by definition does not affect $Y_3$--see~\cite[page 188, second to last equation in the left column]{VinodRX}).

\smallskip
Example~2: Out-of-band two-way-like rate limited output feedback~\cite{suhISIT2012}.
Is this case $Y_1,Y_2$ (the channel outputs at the sources)
are noisy functions of $X_3,X_4$ (the channel inputs from the destinations), and 
$Y_3,Y_4$ (the channel outputs at the destinations) are noisy functions of 
$X_1,X_2$ (the channel inputs from the sources). This implies,
assuming independent noises,  that
\begin{align*}
  &R_1+R_2
\\&\leq I(X_2;Y_4,Y_1| Y_3,X_1,X_3,X_4)+I(X_1,X_2,X_4;Y_3|X_3)
\\&\leq I(X_2;Y_4| Y_3,X_1)+ I(X_1,X_2;Y_3)
\end{align*}
which is formally the same sum-rate bound as in the classical IFC without cooperation~\cite{kramer}--however here the inputs $X_1,X_2$ can be correlated.
The work in~\cite{suhISIT2012}, which was limited to the high-SNR linear deterministic approximation of the Gaussian noise channel, showed that the above sum-rate evaluates to~\cite[eq.(6)]{suhISIT2012}. Notice that the model in~\cite{suhISIT2012} subsumes the one in~\cite{VahidRateLimitedFeedback} and therefore it is not surprising that the above sum-rate bound is the same as~\cite[eq.(22e)]{VahidRateLimitedFeedback}.

\smallskip
{\bf Remark: Extension of Thm.~\ref{theorem:out for general IFC-GF} to the  $K$-CoopIFC with $K>2$.}
Thm.~\ref{theorem:out for general IFC-GF} can be generalized to any number of two-way pairs and any partial sum-rate in the spirit of~\cite{tuniita2010} as outlined in the Appendix.
With this extension, one recovers for example the sum-rate upper bound of~\cite[Thm.4]{tadonZfeedback}.

\smallskip
{\bf Remark: Extension of Thm.~\ref{theorem:out for general IFC-GF} to other channel models.}
We think of Thm.~\ref{theorem:out for general IFC-GF} as a generalization of Kramer's~\cite[Th.1]{kramer}, originally derived for the classical non-cooperative Gaussian IFC. Kramer's idea has been generalized by the author and her collaborators to other interference networks such as: 
(i) the 2-user cognitive IFC~\cite{RiniJ1,RiniJ2}, where the bound is tight for the sum-rate of semi-deterministic channels~\cite{RiniJ1} and tight to within one bit for the Gaussian channel~\cite{RiniJ2},
(ii) the 2-user IFC with a cognitive relay~\cite{RiniITWDubli2010}, and, where the bound is tight for the sum-rate of the linear deterministic approximation of the Gaussian noise channel~\cite{RiniITWDubli2010,DytsoICC2012}, and
(iii) the 3-user cognitive IFC with cumulative message sharing, where the bound is tight for the sum-rate of the linear deterministic approximation of the Gaussian noise channel~\cite{MaamariISIT2012}.

\smallskip
The discussion in the previous paragraph points to a fact observed few times in the past,  that `the same bound seems to apply to different channel models'. For example, the symmetric generalized degrees of freedom for the classical 2-user IFC~\cite{etw} coincides for certain parameters with that of source cooperation~\cite{echoasilomar2011} (which includes as special case the 2-user IFC with output feedback from the source to the intended destination~\cite{SuhTseOutputFeedback}) or for the 2-user cognitive channel~\cite{MaamariISIT2012} (where one source has a priori non-causal message knowledge about the message of the other source). We now try to understand why this is so by analyzing the steps of the proof of Thm.~\ref{theorem:out for general IFC-GF}, in particular we ask whether equalities hold under more general assumptions than those listed for the general cooperative IFC. 

The critical equalities in the derivation of~\eqref{kramer-sr a} are those where we increased conditioning in the entropy terms with positive sign in such a way that the entropy is not reduced; the inequalities are all due to the non-negativity of mutual information or to the fact that conditioning reduces entropy and therefore hold for any channel.
\begin{subequations}
In particular, in~\eqref{eq:kra steps def.enc3} we used the definition of encoding function at node 3, i.e., 
\begin{align}
  &X_{3,t}(Y_3^{t-1}),
\end{align}
while in~\eqref{eq:kra steps def.enc1-34}  we used the definition of encoding functions at nodes 1, 3 and 4. However,
after a more careful inspection of~\eqref{eq:kra steps def.enc1-34}, one realizes that equality holds whenever
\begin{align}
  &X_{1,t}(W_1,Y_1^{t-1},Y_3^{t-1},Y_4^{t-1}), 
\\&X_{4,t}(W_1,Y_1^{t-1},Y_3^{t-1},Y_4^{t-1}),
\end{align}
as one can include in the input definition all variables that appear in the conditioning.
The above condition has the following interpretation: the bound holds even when nodes~1 and~4 are (i) collocated, (ii) have non-causal knowledge/cognition of the message sent by node~1, and (iii) have causal output feedback about the received signal at node~3 (we keep it causal to have a meaningful practical system).
The definition of encoding at node~2 was actually never used; this implies that $X_{2,t}$ can be any function of the messages and of the channel outputs; in particular it may include output feedback and non-causal knowledge/cognition of the messages as follows
\begin{align}
X_{2,t}(W_1,W_2,Y_1^{t-1},Y_2^{t-1},Y_3^{t-1},Y_4^{t-1}).
\end{align}
\label{eq:encoding conditions}
The conditions in~\eqref{eq:encoding conditions} therefore explain why the upper bound in~\eqref{kramer-sr a} holds for different channels, including the cognitive interference channel~\cite{RiniJ2}, the interference channel with output feedback~\cite{SuhTseOutputFeedback},  with generalized feedback~\cite{tuniita2010}, and the two-way-like cooperative channel~\cite{suhISIT2012}.

Note that the upper bound in~\eqref{kramer-sr b}, obtained from the the upper bound in~\eqref{kramer-sr a} by swapping the role of the user pairs, requires imposing the equivalent of the conditions in~\eqref{eq:encoding conditions} obtained by swapping the role of the users. By doing so, we arrive at the following conclusion:  Thm.~\ref{theorem:out for general IFC-GF} holds for memoryless cooperative channels such that
\begin{align*}
  &X_{1,t}(W_1,Y_1^{t-1},Y_3^{t-1},Y_4^{t-1}), \
   X_{2,t}(W_2,Y_2^{t-1},Y_3^{t-1},Y_4^{t-1}), 
\\&X_{3,t}(Y_3^{t-1}), \ X_{4,t}(Y_4^{t-1}),
\end{align*}
that is, for a general cooperative channel in which the `generalized feedback' signals $Y_1,Y_2$ may include any combination of `output feedback signals'  $Y_4,Y_3$. 
\end{subequations}

\smallskip
{\bf Remark: On the tightness of Thm.~\ref{theorem:out for general IFC-GF}.}
Thm.~\ref{theorem:out for general IFC-GF} is not tight in general. However, because it does not contain auxiliary random variables, it can be easily evaluated for many channels of interest, including the Gaussian noise channel. Thm.~\ref{theorem:out for general IFC-GF} is not even tight for the symmetric sum-rate of the classical non-cooperative IFC, for which the novel `weak interference' upper bound discovered by Etkin Tse and Wang~\cite[Thm.1]{etw} is needed. The `ETW-type' upper bound in~\cite[Thm.1]{etw} requires certain invertibility conditions on the channel output functions, in the spirit of~\cite{costaelgamal,telatartse}. Such extensions of the `ETW-type' upper bound have been found in the literature for special types of cooperation. However, the generalization to the general cooperative IFC appears very challenging as remarked in~\cite{suhISIT2012}. 

Another family of bounds for channels with feedback can be obtained by applying Willems' dependance balance idea~\cite{TadonUlukus}. For example in~\cite{TadonUlukus} it was shown that the dependance balance bound can be tighter than the bounds reported in this paper for the Gaussian noise case at small SNR. Therefore, in order to obtain the tightest possible outer bound region, dependance balance bounds need to be considered in general.

\section{The Gaussian Noise Channel}
\label{sec:awgn}

Here we consider the following symmetric version of the channel in~\eqref{eq:awgnch}.
For some $\mathsf{snr}\in\mathbb{R}^+$  parameterize  
\begin{align*}
&\Big[|\mathrm{h}_{\ell,i}|^2 \mathsf{P}_i\Big]_{(\ell,i)\in[1:4]\times[1:4]} = 
\begin{bmatrix}
\star & \mathsf{snr}^{\beta_{\rm s}} & \mathsf{snr}^{\gamma}& \mathsf{snr}^{\gamma\tilde{\alpha}}\\
\mathsf{snr}^{\beta_{\rm s}} & \star & \mathsf{snr}^{\gamma\tilde{\alpha}}& \mathsf{snr}^{\gamma}\\
\mathsf{snr} & \mathsf{snr}^{\alpha} & \star & \mathsf{snr}^{\beta_{\rm d}} \\
\mathsf{snr}^{\alpha} & \mathsf{snr} & \mathsf{snr}^{\beta_{\rm d}} & \star \\
\end{bmatrix},
\\
&\mathsf{C}_1=\mathsf{C}_2=\kappa \log_2(1+\mathsf{snr}),
\\
&\mathsf{C}_3=\mathsf{C}_4=0,
\end{align*}
where $\star$ means that the corresponding value does not affect the capacity region,
and define
\begin{align*}
d_i := \frac{R_i}{\log_2(1+\mathsf{snr})}, \ i\in[1:2].
\end{align*}
The {\em generalized degrees of freedom region} is the upper convex envelope of the set of achievable $(d_{1},d_{2})$ for some $\mathsf{snr}\geq 0$. The {\em symmetric generalized degrees of freedom} is
\[
d:= \max\{d_{1}+d_{2}\}/2,
\]
where the maximum is over the generalized degrees of freedom region.
\begin{subequations}
The generalized degrees of freedom of the symmetric Gaussian noise channel satisfies:
from the cut-set bound in Thm.~\ref{theorem:out cutset}
\begin{align}
   d &\leq \max\{1,\alpha,\beta_{\rm d}\}
\\ d &\leq \max\{\beta_{\rm s}+\beta_{\rm d},1+\gamma\} + \Delta_1, 
\\ \Delta_1&:=\max H(f_2|X_2,X_3,V_2,V_3,Y_3) \leq \kappa, \label{eq:f2 delta1} 
\\   d &\leq \max\{\beta_{\rm s},1,\alpha\}  + \Delta_2,
\\ \Delta_2&:=\max H(f_2|X_2,X_3,X_4,V_2,V_3,V_4,Y_3,Y_4) \leq \kappa, \label{eq:f2 delta2} 
\\2d &\leq 2\max\{1,\alpha\}: \alpha \not= 1, \label{eq:ultimate inband 1}  
\\2d &\leq 1  :\  \alpha = 1 \label{eq:ultimate inband 2}
\end{align}
and from Thm.~\ref{theorem:out for general IFC-GF}
\begin{align}
  2d &\leq \max\{1,\alpha,\beta_{\rm d}\}+[\max\{\beta_{\rm s},1\}-\alpha]^+ + \Delta_2,
\end{align}
\label{allbounds}
where $f_2$ in~\eqref{eq:f2 delta1} and~\eqref{eq:f2 delta2} is the out-of-band part of the generalized feedback signal received at node~2 as defined by~\eqref{eq:awgnch}; note that here we assumed that $f_1$ is obtained from $f_2$ by swapping the role of the users. Note that in general $\Delta_2\leq \Delta_1$.
The ultimate limit of cooperation is given by~\eqref{eq:ultimate inband 1}-\eqref{eq:ultimate inband 2}, which corresponds to the capacity of a $2\times 2$ MIMO point-to-point channel; the discontinuity at $\alpha=1$ is due to the fact that at $\alpha=1$ the $2\times 2$ MIMO channel matrix becomes rank-deficient and therefor there is a loos in degrees of freedom.  Interestingly, the upper bound in~\eqref{allbounds} does not depend on the parameter $\tilde{\alpha}$.
\end{subequations}

\begin{figure}
\centering
\vspace*{-3cm}
\includegraphics[width=10cm]{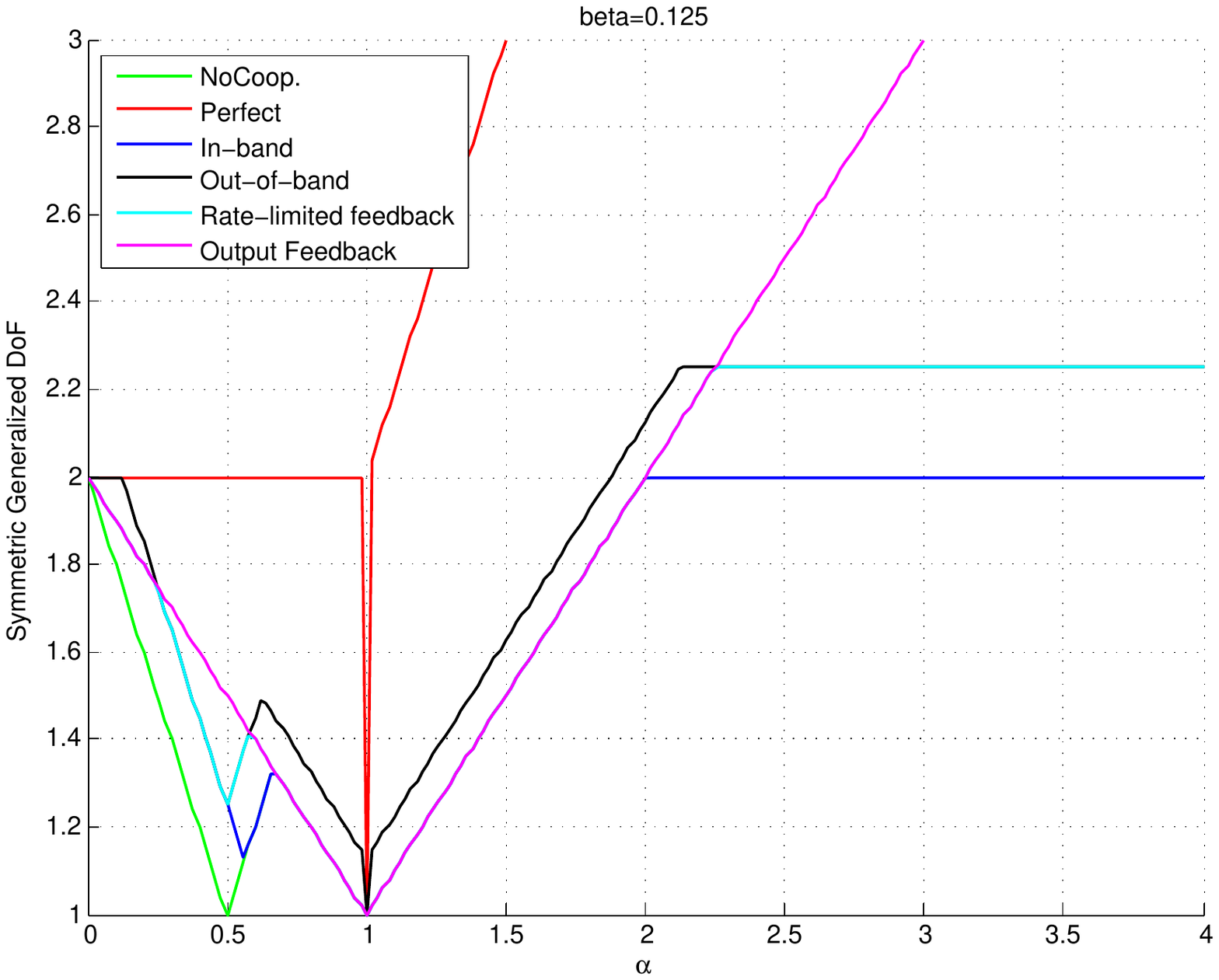}
\vspace*{-4cm}
\caption{Symmetric generalized degrees of freedom vs. $\alpha$ for $\beta=0.125$.}
\label{fig:very small beta}
%
%
\vspace*{-3cm}
\includegraphics[width=10cm]{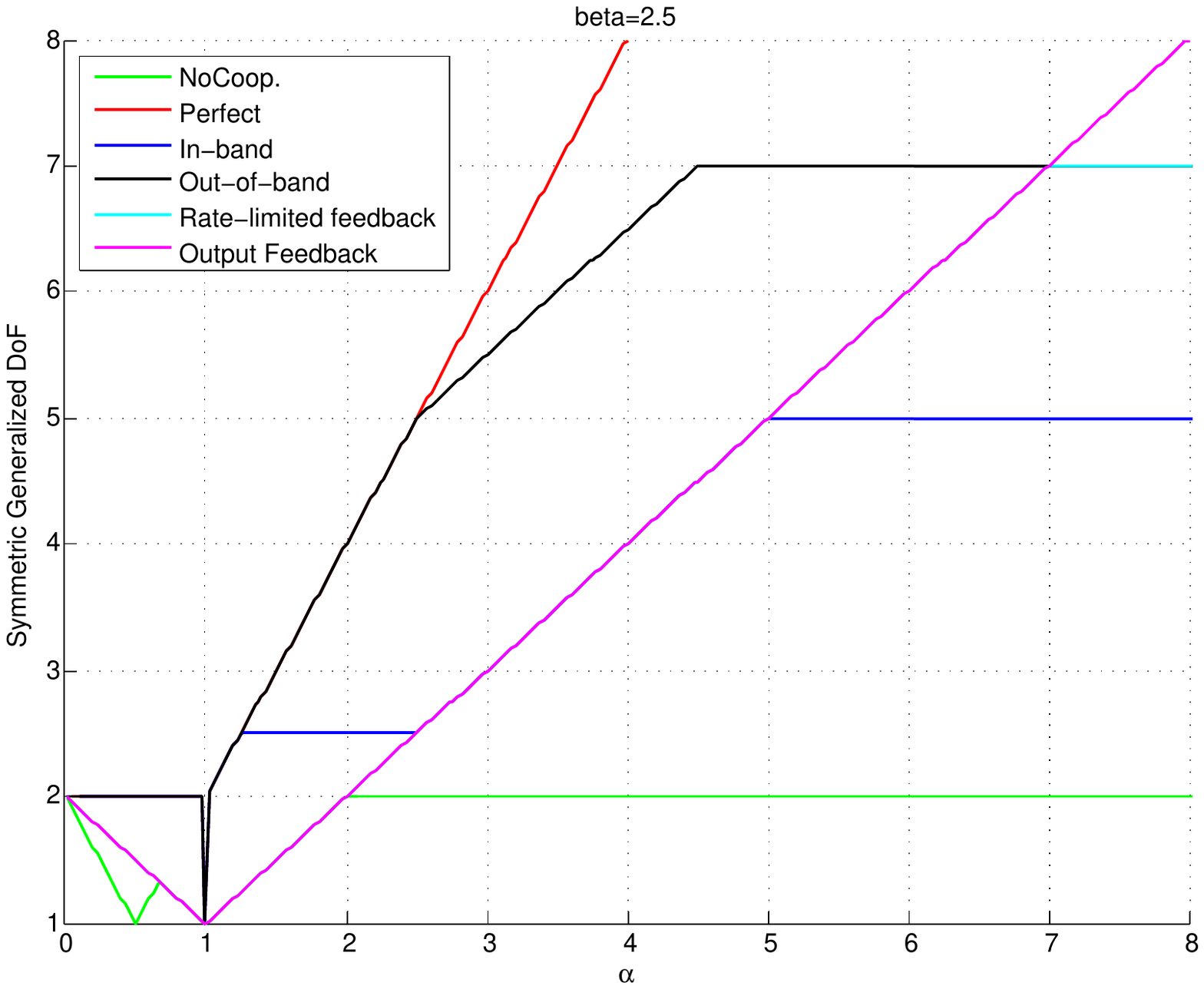}
\vspace*{-4cm}
\caption{Symmetric generalized degrees of freedom vs $\alpha$ for $\beta=2.5$.}
\label{fig:large beta}
\end{figure}

The upper bounds in~\eqref{allbounds} allow to compare different types of cooperation for a given interference level $\alpha$.
Interesting conclusions can be drawn from this comparison. Figs.~\ref{fig:very small beta} and~\ref{fig:large beta} show, as a function of the interference level $\alpha$ and for a fixed $\beta$ (whose value is indicated in the caption and whose meaning depends on the mode of cooperation) the total symmetric generalized degrees of freedom ($2d$). If the upper bound in~\eqref{allbounds} is not tight to within a constant gap we indicate the regime where a generalization of~\cite[Th.1]{etw} is tight to within a constant gap. We have the following modes of cooperation:
\begin{itemize}
\item
No-cooperation~\cite{etw}: $\beta_{\rm s}=\beta_{\rm d}=\gamma=\tilde{\alpha}=\kappa=0$; in this case the symmetric generalized degrees of freedom is given by the so-called W-curve $d = \min\{\min\{1-\alpha,\alpha\}, \max\{1-\alpha/2,\alpha/2\}, 1\}:=\mathsf{W}(\alpha)$;
\item
In-band source cooperation~\cite{VinodTX,echoasilomar2011}: $\beta_{\rm s}=\beta, \ \beta_{\rm d}=\gamma=\tilde{\alpha}=\kappa=0$;~\eqref{allbounds} is tight except for $\alpha<2/3, \beta<\alpha/2$; we note that in-band destination cooperation achieves the same symmetric generalized degrees of freedom as in-band source cooperation~\cite{VinodRX};
\item
Out-of-band source cooperation or conferencing encoders~\cite{ihunagtseTX}: $\kappa=\beta, \ \beta_{\rm s}=\beta_{\rm d}=\gamma=\tilde{\alpha}=0$ and $\Delta_1=\Delta_2=\kappa$;~\eqref{allbounds} is tight except for $\alpha<2/3, \beta<\min\{\alpha,2-3\alpha\}$; we note that out-of-band destination cooperation achieves the same symmetric generalized degrees of freedom as out-of-band source cooperation~\cite{ihunagtseTX};
\item
Output feedback~\cite{SuhTseOutputFeedback}: $\kappa=\infty, \ \beta_{\rm s}=\beta_{\rm d}=\gamma=\tilde{\alpha}=\kappa=0$; in this case the symmetric generalized degrees of freedom is given by the so-called V-curve $d = \max\{1-\alpha/2,\alpha/2\}:=\mathsf{V}(\alpha)$;
\item
Out-of-band rate-limited feedback~\cite{VahidRateLimitedFeedback}: $\kappa=\beta, \ \beta_{\rm d}=\beta_{\rm s}=\gamma=\tilde{\alpha}=0$  and $\Delta_2=0 < \Delta_1=\kappa$; in this case the symmetric generalized degrees of freedom is given $d=\min\{\mathsf{V}(\alpha), \mathsf{W}(\alpha)+\beta\}$, that is, the bound in~\eqref{allbounds} is tight except for $\alpha<2/3, \beta<\min\{\alpha/2,[2-3\alpha]^+/2\}$; we note that rate-limited feedback performs as output feedback~\cite{SuhTseOutputFeedback} in terms of symmetric generalized degrees of freedom except for $\beta<\min\{\alpha/2,[2-3\alpha]^+/2\}$ and for $\beta>[2-\alpha]^+/2$;
\item
Ultimate limit of cooperation~\cite{book:cover_thomas:it}: $\beta_{\rm s}=\beta_{\rm d}=\infty$ (here $\gamma,\tilde{\alpha}$ and $\kappa$ do not matter; in this case the symmetric generalized degrees of freedom is given by $d=\max\{1,\alpha\}$ except at $\alpha=1$ where $d=1/2$.
\end{itemize}
We do not report the symmetric generalized degrees of freedom for the two-way-like cooperation~\cite{suhISIT2012}
($\kappa>0, \ \beta_{\rm d}=\beta_{\rm s}=\gamma=\tilde{\alpha}=0$ and $\Delta_2=0 < \Delta_1=\kappa$) because
it is not known at present whether~\eqref{allbounds} is tight to within a constant gap.

By observing Figs.~\ref{fig:very small beta} and~\ref{fig:large beta} we make the following remarks.
\begin{itemize}
\item
In-band source cooperation.
It does not require extra resources in terms of power or bandwidth with respect to the no-cooperation case.
It achieves the same symmetric generalized degrees of freedom as output feedback except for a subset of $\alpha\in[1/2,2/3]$ and for $\alpha>2$--see remark in Section~\ref{sec:outer}.
It does not improve on the no-cooperation case for a subset of $\alpha\in[2/3,2]$--see remark in Section~\ref{sec:outer}.
It achieves the ultimate limit of cooperation for $\beta>\max\{1+\alpha,2\alpha\}$.

\item
Out-of-band source cooperation.
It is always better than in-band source cooperation because of the extra dedicated channels for conferencing (through which a source can send data that differ from what it sends on the main channel and therefore that data is always be useful for the cooperating source). We argue that a more fair comparison should be made between in-band and out-of band cooperation in which the total transmit power and the total bandwidth in the two models are kept the same.

\item
Output feedback cooperation.
It achieves the same performance as in-band source cooperation with $\beta=\alpha/2$; this means that there is no need to deploy dedicated channels for conferencing if $|\mathsf{h}_{1,2}|^2 = |\mathsf{h}_{2,1}|^2 = \mathsf{snr}^\beta \geq \sqrt{\mathsf{snr}^\alpha}$ with $|\mathsf{h}_{4,2}|^2 = |\mathsf{h}_{3,1}|^2 = \mathsf{snr}$ and $|\mathsf{h}_{4,1}|^2 = |\mathsf{h}_{3,2}|^2 = \mathsf{snr}^\alpha$.  It never achieves the ultimate limit of cooperation.

\item
Out-of-band rate-limited feedback.
Its performance are upper bounded by the case of output feedback cooperation and the two actually coincide for $\alpha<2(1+\beta)$.
Note that output feedback cooperation is also an upper bound for the case of two-way-like cooperation because; this can be seen by using the results on all possible configuration of output feedback from a source to a destination~\cite{melda}.
\end{itemize}

\section{Conclusions}
\label{sec:conc}
In this paper we developed a sum-rate upper bound for the general memoryless cooperative interference channel that generalizes all those bounds known for special types of cooperation that reduce to Kremer's bound in the case of no-cooperation. It is part of ongoing work to develop a general framework for those sum-rate bounds that reduce to Etkin Tse and Wang's novel upper bound in the case of no-cooperation.

\section*{Acknowledgment}
The work was partially funded by NSF under award number 0643954.
The contents of this article are solely the responsibility of the author and
do not necessarily represent the official views of the NSF.

This work was possible thanks to the generous support of Telecom-ParisTech, Paris France,
while the author was on a sabbatical leave at the same institution.
The author would like to thank Dr. M. Wigger for insightful discussions on the role of feedback in memoryless networks.

%

{

}

\section*{Appendix}
By Fano's inequality we have
\[
\max_{i\in[1:K]}\{H(W_{i}|Y_{i+K}^N)\}\leq N\epsilon_N,
\]
for some $\epsilon_N \to 0$ as $N\to\infty$, which implies
\begin{align*}
  N (R_{i}-\epsilon_N) \leq I(W_{i}; Y_{i+K}^N), \ i\in[1:K].
\end{align*}
For~\eqref{kramer-sr a} (the bound in~\eqref{kramer-sr b} follows by reversing the role of the users) we have:
\begin{subequations}
\begin{align}
  &N(R_1+R_2-2\epsilon_N)
\leq I(W_1; Y_3^N)
   + I(W_2; Y_4^N)
\nonumber\\&
\leq I(W_1; Y_3^N)
   + I(W_2; Y_4^N, \ Y_1^N,Y_3^N,W_1)
\nonumber\\&
=    I(W_1; Y_3^N)
   + I(W_2; Y_4^N, \ Y_1^N,Y_3^N|W_1)
\nonumber\\&= H(Y_3^N)+H(Y_4^N,Y_1^N|W_1,Y_3^N)
- H(Y_1^N,Y_3^N,Y_4^N|W_1,W_2)
\nonumber\\&=\sum_{t}
    \label{eq:kra steps def.enc3}
     H(Y_{3,t}|         Y_3^{t-1},{\blue X_3^t})
\\&+ \label{eq:kra steps def.enc1-34}
     H(Y_{4,t},Y_{1,t}| W_1,Y_3^N,Y_1^{t-1},Y_4^{t-1},{\blue X_3^N,X_1^t,X_4^t})
\\&-
    H(Y_{1,t},Y_{3,t},Y_{4,t}| W_1,W_2,Y_1^{t-1},Y_3^{t-1},Y_4^{t-1})
\nonumber\\&\leq\sum_{t}
    \label{eq:kra steps cond.red.entropy 1}
    H(Y_{3,t}| X_{3,t})
  + H(Y_{4,t},Y_{1,t}| Y_{3,t},X_{1,t},X_{3,t},X_{4,t})
\\&-
    H(Y_{1,t},Y_{3,t},Y_{4,t}| W_1,W_2,Y_1^{t-1},Y_3^{t-1},Y_4^{t-1})
\nonumber\\&\leq\sum_{t}
    H(Y_{3,t}| X_{3,t})
  + H(Y_{4,t},Y_{1,t}| Y_{3,t},X_{1,t},X_{3,t},X_{4,t})
\\&-
    \label{eq:kra steps add cond. 1}
    H(Y_{1,t},Y_{3,t},Y_{4,t}| W_1,W_2,Y_1^{t-1},Y_3^{t-1},Y_4^{t-1},{\blue X_1^t,X_2^t,X_3^t,X_4^t}) 
\nonumber\\&=\sum_{t}
    H(Y_{3,t}| X_{3,t})
  + H(Y_{4,t},Y_{1,t}| Y_{3,t},X_{1,t},X_{3,t},X_{4,t})
\\&-
    \label{eq:kra steps memoryless 1}
    H(Y_{1,t},Y_{3,t},Y_{4,t}| X_{1,t},X_{2,t},X_{3,t},X_{4,t})
\\&=\sum_{t}
    I(Y_{3,t};X_{1,t},X_{2,t},X_{4,t}|X_{3,t})
\nonumber\\&+
    I(Y_{4,t},Y_{1,t};X_{2,t}|X_{1,t},Y_{3,t},X_{3,t},X_{4,t}),
\nonumber
\end{align}
\end{subequations}
where
in~\eqref{eq:kra steps def.enc3} and~\eqref{eq:kra steps def.enc1-34} 
we used the definition of encoding functions (terms in blue),
in~\eqref{eq:kra steps cond.red.entropy 1} (and also in~\eqref{eq:kra steps add cond. 1}) we used ``conditioning reduces entropy'', and
in~\eqref{eq:kra steps memoryless 1} the fact that the channel is memoryless.

We next outline how Thm.~\ref{theorem:out for general IFC-GF} can be extended to any number of two-way pairs and any partial sum-rate. We exemplify our approach for the case $K=4$ and the partial sum-rate $R_1+R_2+R_3$; generalization to other $K$ or other sum-rates is straightforward. We have
\begin{subequations}
\begin{align*}
  &N(R_1+R_2+R_3-3\epsilon_N)
\\&\leq
     I(W_1; Y_{1+K}^N)
   + I(W_2; Y_{2+K}^N)
   + I(W_3; Y_{3+K}^N)
\\&\leq
     I(W_1; Y_{1+K}^N, \ [W_4,Y_{4}^N]) 
\\&+ I(W_2; Y_{2+K}^N, \ [Y_{1+K}^N,W_1,Y_{1}^N], \ [W_4,Y_{4}^N])  
\\&+ I(W_3; Y_{3+K}^N, \ [Y_{1+K}^N,W_1,Y_{1}^N], \ [Y_{2+K}^N,W_2,Y_{2}^N], \ [W_4,Y_{4}^N])
%
\\&=
     H(Y_{1+K}^N, Y_{4}^N | W_4)
\\&+ H(Y_{2+K}^N,Y_{1}^N| W_1,W_4, Y_{1+K}^N,Y_{4}^N) 
\\&+ H(Y_{3+K}^N,Y_{2}^N| W_1,W_2,W_4,   Y_{2+K}^N,Y_{1}^N,Y_{1+K}^N,Y_{4}^N)
\\&- H(Y_{3+K}^N,Y_{2}^N,   Y_{2+K}^N,Y_{1}^N,Y_{1+K}^N,Y_{4}^N| W_1,W_2,W_3,W_4)
\\&\leq \sum_{t}
     H(Y_{1+K,t},Y_{4,t}| X_{4,t})
\\&+ H(Y_{2+K,t},Y_{1,t}| X_{1,t},X_{4,t}, Y_{1+K,t},Y_{4,t})
\\&+ H(Y_{3+K,t},Y_{2,t}| X_{1,t},X_{2,t},X_{4,t},   Y_{2+K,t},Y_{1,t},Y_{1+K,t},Y_{4,t})
\\&- H(Y_{3+K,t},Y_{2,t},   Y_{2+K,t},Y_{1,t},Y_{1+K,t},Y_{4,t}| X_{1,t},X_{2,t},X_{3,t},X_{4,t})
%
\\&\leq \sum_{t}
     I(X_{1,t},X_{2,t},X_{3,t}; Y_{1+K,t},Y_{4,t}| X_{4,t})
\\&+ I(        X_{2,t},X_{3,t}; Y_{2+K,t},Y_{1,t}| X_{1,t},X_{4,t},                             Y_{1+K,t},Y_{4,t})
\\&+ I(                X_{3,t}; Y_{3+K,t},Y_{2,t}| X_{1,t},X_{2,t},X_{4,t},   Y_{2+K,t},Y_{1,t},Y_{1+K,t},Y_{4,t}).
\end{align*}
\end{subequations}

\end{document}